\documentclass[useAMS,usenatbib]{mn2e}

\usepackage{epsfig}
\usepackage{textcomp}
\usepackage{longtable}
\usepackage{lscape}

\newcommand{\xmm} {{\it XMM-Newton}}

\newcommand{\cmsq} {cm$^{-2}$}
\newcommand{\nh} {$N_{\rm{H}}$}
\newcommand{\lx} {$L_{\rm{X}}$}

\newcommand{\mic}{{${\umu}$m}}
\newcommand{\oiii}{{\rm{[O\,\sc{iii}]}}}
\newcommand{\hii}{{\rm{H\,\sc{ii}}}}

\title[]{X-ray colour-colour selection for heavily absorbed AGN}

\author[M. Brightman and K. Nandra]{Murray Brightman$^{1}\thanks{E-mail: mbright@mpe.mpg.de}$ and Kirpal Nandra$^{1}$\\
$^{1}$Max-Planck-Institut f\"{u}r extraterrestrische Physik, Giessenbachstrasse 1, D-85748, Garching bei M\"{u}nchen, Germany\\}

\begin{document}

\date{Accepted 0000 December 00. Received 0000 December 00; in original form 0000 October 00}

\pagerange{\pageref{firstpage}--\pageref{lastpage}} \pubyear{0000}

\maketitle

\label{firstpage}

\begin{abstract}
We present a method for the identification of heavily absorbed AGN (\nh$>10^{23}$ \cmsq) from X-ray photometric data. We do this using a set of \xmm\ reference spectra of local galaxies for which we have accurate \nh\ information, as described in Brightman \& Nandra. The technique uses two rest-frame hardness ratios which are optimised for this task, which we designate HR1 (2-4/1-2 keV) and HR2 (4-16/2-4 keV). The selection method exploits the fact that while obscured AGN appear hard in HR2 due to absorption of the instrinsic source flux below $\sim$ 4 keV, they appear soft in HR1 due to excess emission originating from scattered source light, thermal emission, or host galaxy emission. Such emission is ubiquitous in low redshift samples. The technique offers a very simple and straightfoward way of estimating the fraction of obscured AGN in samples with relatively low signal-to-noise ratio in the X-ray band. We apply this technique to a moderate redshift ($z\sim1$) sample of AGN from the {\it Chandra} Deep Field North, {\bf finding that 61\% of this sample has \nh$>10^{23}$ \cmsq}. A clear and robust conclusion from our analysis, is that in deep surveys the vast majority of sources do not show hardness ratios consistent with a simple absorbed power law. The ubiquity of complex spectra in turn shows that simple hardness ratio analysis will not yield reliable obscuration estimates, justifying the more complex colour-colour analysis described in this paper. {\bf While this method does very well at separating sources with \nh$>10^{23}$ \cmsq\ from sources with lower \nh, only X-ray spectroscopy can identify Compton thick sources, through the detection of the Fe K$\alpha$ line. This will be made possible with the high throughput X-ray spectral capabilities of {\it ATHENA}}
\end{abstract}


\section{Introduction}

The nature and amount of obscuration in active galactic nuclei (AGN) is an important issue for a range of astrophysical questions, such as unification of AGN \citep[eg.][]{antonucci93} and the role of AGN in galaxy evolution \citep{hopkins06}. Heavily absorbed AGN can be identified using their X-ray spectra, via direct measurement of the line of sight column density. At the highest column densities ($N_{\rm H}>10^{24}$ cm$^{-2}$, where even the direct X-rays are suppressed, obscured AGN can still be identified via the presence of X-ray reflection, with the spectrum consisting of a flat X-ray continuum accompanied by a high equivalent width Fe K$\alpha$ line \citep*{matt96_2}. The measurement of column densities or identification of reflection features requires good quality X-ray spectra with enough counts to carry out spectral fitting and to reject alternative hypotheses. This is challenging, even in the local universe, as the flux suppression in the 2-10 keV band is significant once the \nh\ reaches a few $10^{23}$ \cmsq\ \citep{brightman11}. Many alternative methods of identifying heavily absorbed sources, where the X-ray spectrum itself is not good enough to do so, have been presented in the literature. These often use the ratio of the observed X-ray flux to the flux at some wavelength at which the effect of the obscuration is assumed to be small, and which should therefore be representative of the AGN bolometric flux. Examples of these assumed ``isotropic indicators'' are the MIR, \oiii\ line or {\rm{[Ne\,\sc{v}]}} line \citep{dstanic09, lamassa10, gilli10}. Obscured AGN may be found via weak observed X-ray emission with respect to the isotropic indicator, with the X-ray flux suppression attributed to absorption. A popular technique which uses this method was presented by \cite{bassani99} and uses $F_{\rm X}/F_{\rm [OIII]}$ in conjunction with the Fe K$\alpha$ line equivalent width as proxies for \nh, though $F_{\rm X}/F_{\rm [OIII]}$ has henceforth often been used alone. The use of isotropic indicators are complicated when the isotropic indicator has contributions from non nuclear sources (e.g. star-formation processes contribute to both MIR and \oiii\ fluxes), or where the source is intrinsically X-ray weak. Furthermore, techniques have even been developed that use data entirely outside of the X-ray band to detect X-ray obscuration \citep[e.g.][]{daddi07, fiore08}. Arguably though, the best place to judge X-ray absorption, is in the X-ray band alone. This is sometimes done using `hardness ratios', which is a model independent technique, and utilises X-ray photometry. Traditionally, hardness ratios are calculated using observed counts, often in the standard soft (0.5-2.0 keV) and hard (2.0-10.0 keV) bands. Sources with large hardness ratio are assumed absorbed, while sources with low ratios assumed unabsorbed. However this method is very crude and may not exclusively separate the obscured sources from the unobscured sources if spectral complexity is present. At other wavelengths, colour-colour selection is an effective method of identifying sources populations based on photometry. Here, we seek a novel approach to the identification of X-ray absorption from X-ray photometric data, using colour-colour selection, formed from X-ray hardness ratios. {\bf While work using a combination of hardness ratios to infer absorption has been presented previously, for example results by \cite{dellaceca99} using {\it ASCA} data}, we seek to improve on these using knowledge from local samples with data of high spectral quality. This better understanding of hardness ratios can then be applied in cases with much poorer data quality, such as in deep surveys of the distant Universe, and help elucidate the observed relationships seen between obscuration, luminosity and redshift \citep[eg.][Brightman \& Ueda, in preparation]{ueda03, hasinger08, brightman11b}.

\section{X-ray data}

For the task of finding an effective colour-colour selection technique for obscured AGN, we utilise a set of 126 \xmm\ spectra of local galaxies, for which we have carried out a detailed spectral analysis of in \citet[][BN11]{brightman11}. The parent sample of this set was the extended {\it IRAS} 12 micron galaxy sample of \cite{rush93}. The mid-IR selection of this sample should ensure relatively little bias against obscured objects \citep{spinoglio89}. The X-ray spectra of these galaxies were fitted in a systematic way, starting with a simple power-law, and adding additional model components, such as absorption or reflection when statistically required. BN11 also applied new Monte-Carlo based models for modelling heavily absorbed sources. Compton thick sources were identified either with a direct measurement of the line of sight \nh, or from the presence of a reflection dominated spectrum, often accompanied by strong Fe K$\alpha$ emission. As such, we have accurate \nh\ information for these galaxies {\bf from \xmm}. As the sample is local (z$<0.1$), we treat the X-ray spectral data as rest-frame. The sample consists of all galaxy types, from Seyfert galaxies to \hii\ and LINER galaxies, with a wide range of \lx, \nh\ and a high fraction of absorbed and Compton thick sources.  For a more detailed description of the X-ray spectral analysis, the X-ray spectral properties of the sample, and the new Monte-Carlo models, see BN11. For the optical activity classification and 12 \mic\ data, see \cite{brightman11b}. 

\begin{figure}
 \includegraphics[width=80mm]{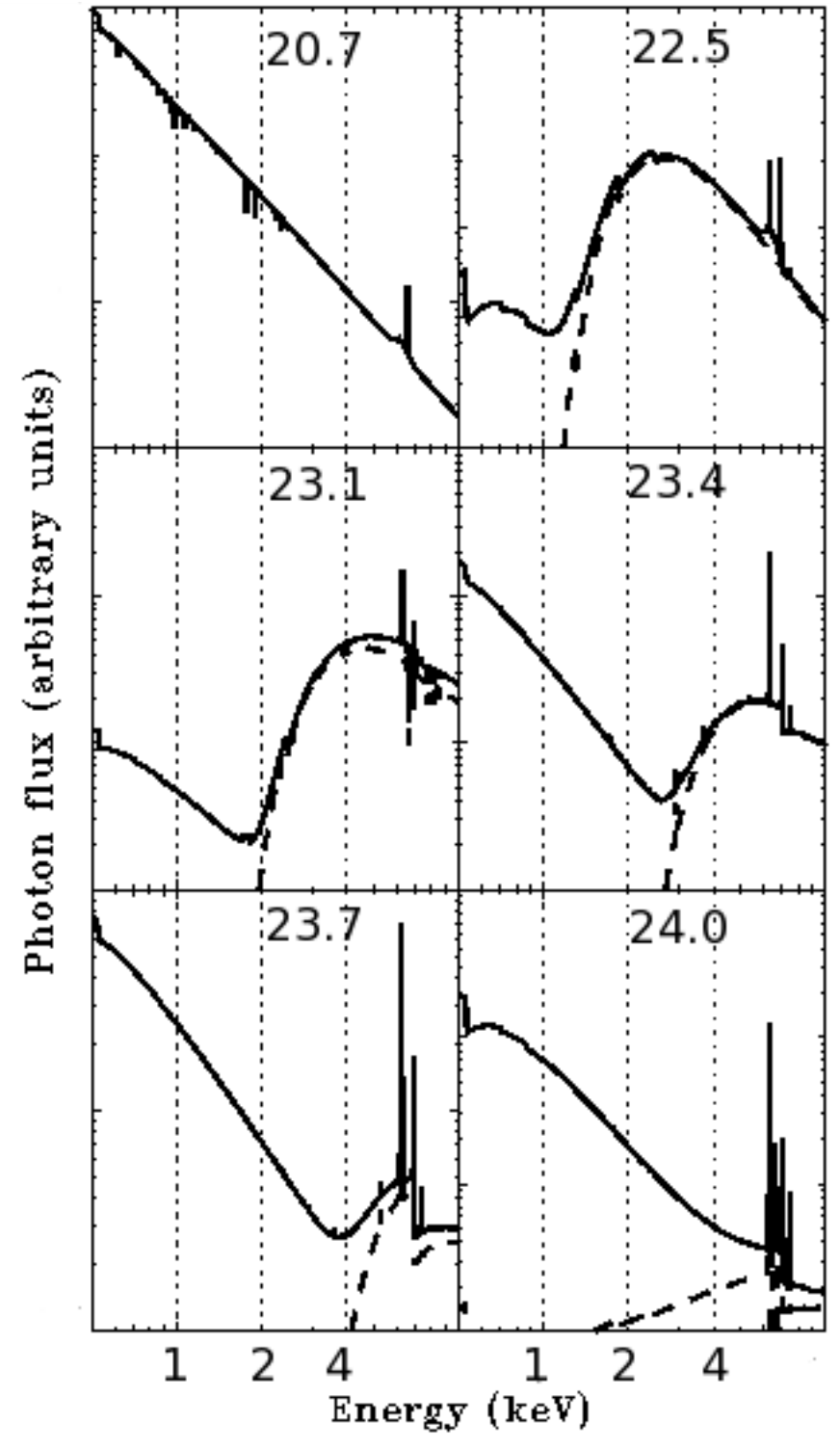}
 \caption{An example set of six of our reference model spectra, for UGC 545, NGC 5506, NGC 262, NGC 2655, NGC 17 and NGC 3079 (top left to bottom right). These show a range in \nh\ values (log$_{10}$(\nh/\cmsq) printed in the corner). The dashed lines show the intrinsic source component, and the dotted lines mark the boundaries of the bands that we use for the calculation of hardness ratios.}
 \label{exspec_fig}
\end{figure}

We use a physically and empirically motivated choice of X-ray bands for our colour-colour plots, with consideration for the Fe K$\alpha$ line and the expected energy of the absorption turn over in heavily obscured AGN. In Fig \ref{exspec_fig} we present a selection of our reference \xmm\ model spectra, over a range in \nh\ values. It can be seen from these that for \nh$>10^{23}$ \cmsq\ there is little direct AGN flux observed below 2 keV, and that emission below 2 keV is dominated by a soft-excess component, which can be attributed to scattered AGN emission, or emission from the host galaxy. As the line of sight \nh\ approaches $10^{24}$ \cmsq, this galactic/scattered emission can dominate even up to 4 keV in the rest frame, and the spectral characteristics of the heavily obscured AGN is seen only above 4 keV. This can be a reflection dominated spectrum, characterised by its flat shape, or a transmitted spectrum with high measured \nh. Both of these are usually accompanied with iron lines of high EW. Based on these physical motivations, we choose our energy band selection to be 1$<$E$<$2 keV (1), 2$<$E$<$4 keV (2) and 4$<$E$<$16 keV (3). To analyse the colours, we then extract model photon fluxes from our set of reference spectra. {\bf While the 4-16 keV data include an extrapolation out of the \xmm\ band, from 10-16 keV, the spectra are well modelled and the \nh\ well defined due to high quality data below 10 keV, making uncertainties in the extrapolation minimal.} We use fluxes rather than count data so that the selection technique is independent of the telescope or instrument used. We calculate hardness ratios based on these fluxes, where HR1=$\frac{F(band 2) - F(band 1)}{F(band 2) + F(band 1)}$ and HR2=$\frac{F(band 3) - F(band 2)}{F(band 3) + F(band 2)}$

\section{Colour-colour selection for heavily obscured AGN}

\begin{figure*}

\includegraphics[width=150mm]{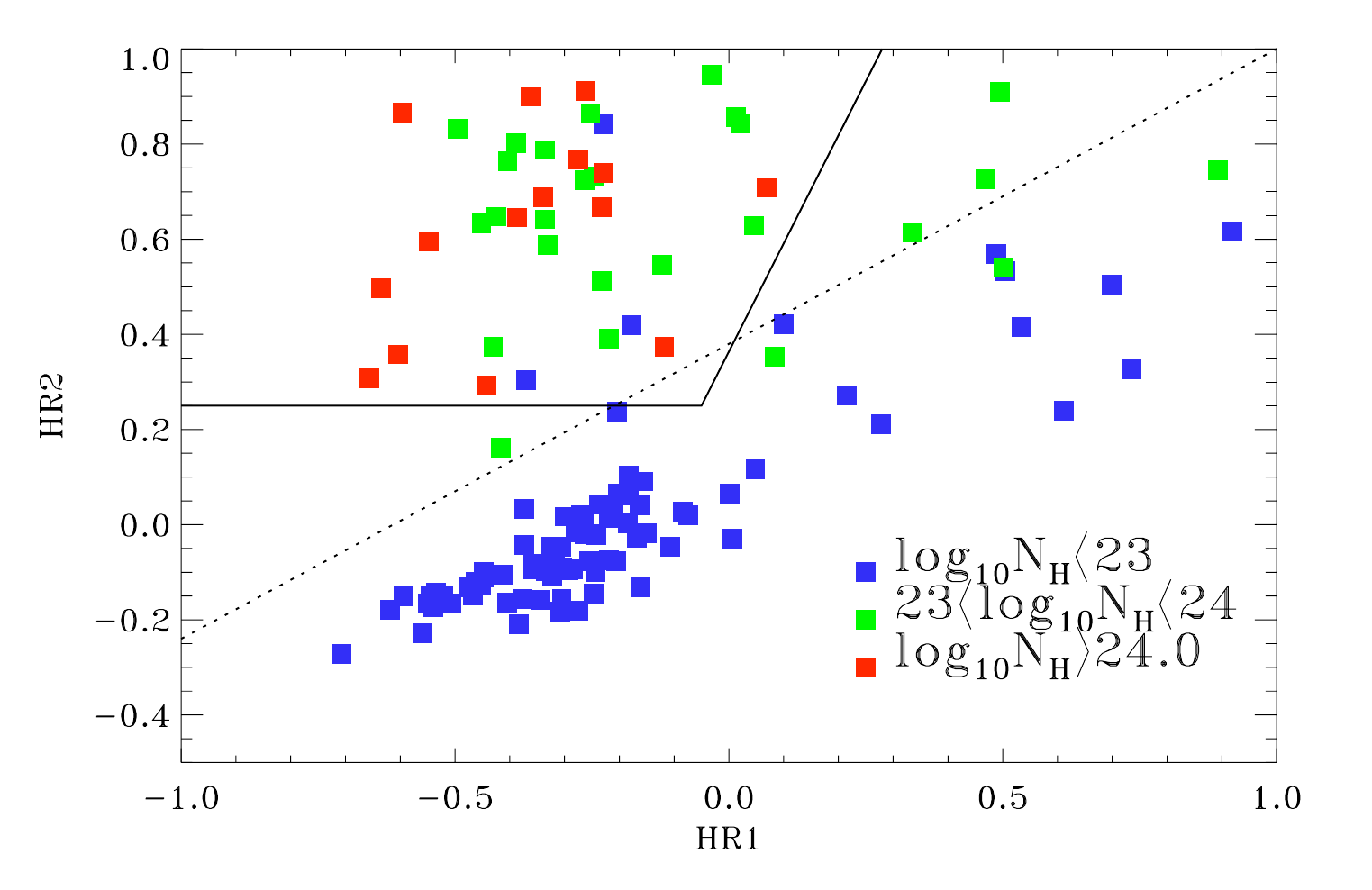}
 \caption{Our colour-colour plot using the hardness ratios HR1 (2-4/1-2 keV) and HR2 (4-16/2-4 keV), calculated from model photon fluxes extracted from our set of 126 \xmm\ reference spectra. The colour of each data point corresponds to the value of the measured \nh\ in that spectrum. The dotted line and the solid wedge are proposed selection regions for heavily obscured AGN, as described in the text. }
 \label{theplot}
\end{figure*}

In Fig. \ref{theplot} we present a colour-colour plot which uses our hardness ratios using our reference spectra data. On it, we also show the \nh\ of each source, represented by colour. There is a cluster of sources at HR1$\simeq$-0.25 and HR2$\simeq$0.0. These are objects which show little absorption above 1 keV. Both ratios harden as the \nh\ increases until $\sim10^{23}$ \cmsq. At this point, the soft-excess emission begins to dominate below 4 keV and hence HR1 softens, while HR2 remains hard. It can be seen that the heavily absorbed sources (\nh$>10^{23}$ \cmsq) are separated well from the others using these hardness ratios in combination, being hard in HR2, while soft in HR1. We therefore plot two proposed selection criteria. First (1), the dotted line selects almost all (38/41=93\%) heavily absorbed sources, with only a few (3/41=7\%) contaminants with a lower \nh. Secondly (2), the solid wedge selects all (16/16=100\%) Compton thick sources. This region is contaminated by 20/36 (56\%) sources with \nh$<10^{24}$ \cmsq, though only by 3/36 (8\%) sources with \nh$<10^{23}$ \cmsq. The selection criteria are as follows. For \nh$>10^{23}$ \cmsq,
 
\begin{equation}
(1)\; {\rm HR2 > 0.62 \times HR1 + 0.38}
\end{equation}
and for Compton thick AGN
\begin{equation}
(2)\; {\rm HR2 > 0.25\; AND\; HR2 < 2.28 \times HR1 + 0.36}
\end{equation}

Calculation of the rest-frame hardness ratios is in practice very straightforward. For example, counts or count rates can be converted to flux to be used in this selection technique using the HEASARC Portable Interactive Mulit-Mission Software ({\sc pimms})\footnote{an online calculator can be found here https://heasarc.gsfc.nasa.gov/Tools/w3pimms.html}. This requires the assumption of a model. For deep X-ray fields, the use of a $\Gamma=1.4$ power-law with no absorption, roughly consistent with the cosmic X-ray background, is appropriate.

\begin{figure}

\includegraphics[width=90mm]{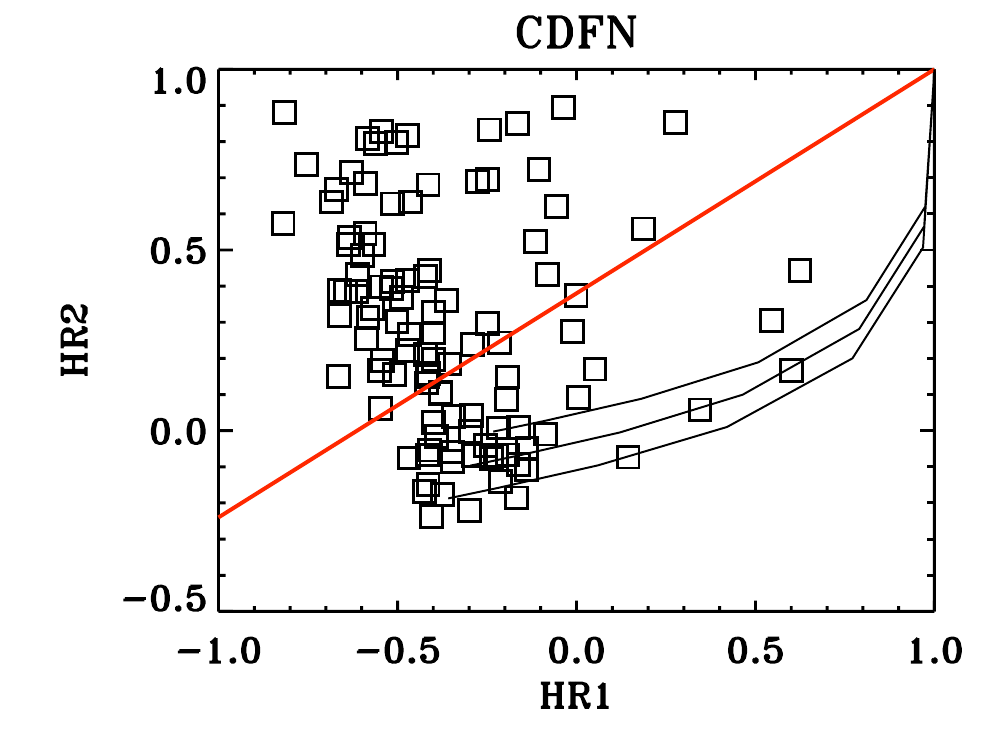}
 \caption{Our hardness ratio diagram applied to data from the CDFN. The solid red line is the selection criterion for \nh$>10^{23}$ \cmsq\ sources, described in the text. The solid lines black lines show the tracks of an absorbed power-law with (top to bottom) $\Gamma$=1.7, 1.9 and 2.1, from \nh=$10^{20}$-$10^{24}$ \cmsq, progressing from left to right. }
 \label{fig_xrshr}
\end{figure}

We can then use this new identification technique to make inferences about the heavily obscured AGN population from currently available X-ray survey data. Here we apply the scheme to the {\it Chandra} Deep Field North survey (CDFN). \cite{alexander03} present photometric data from this survey in the 0.5-1 keV, 1-2 keV and 2-8 keV bands. These correspond exactly to our band selection for $z=1$ sources. Redshift data for this catalogue have been provided by \cite{trouille08}, including photometric redshifts. Using these, we select sources with $0.8<z<1.2$.  The hardness ratio plot for this survey is presented in Fig. \ref{fig_xrshr}. Here we also show tracks of a simple absorbed power-law, calculated using the {\it zwabs*power-law} models in {\sc xspec} with $\Gamma$=1.7, 1.9 and 2.1, from \nh=$10^{20}$-$10^{24}$ \cmsq.

We find that for the 107 $0.8<z<1.2$ CDFN sources we have selected, 65 (61\%) lie above the selection line. We can also use these hardness ratios to infer information about the spectral complexity present in these sources. Only 39 (36\%) of the CDFN sources we study here show spectra consistent with a simple absorbed power-law, where the majority of sources present more complex spectra, due to soft excess emission. 

\section{discussion and summary}

\begin{figure}
 \includegraphics[width=90mm]{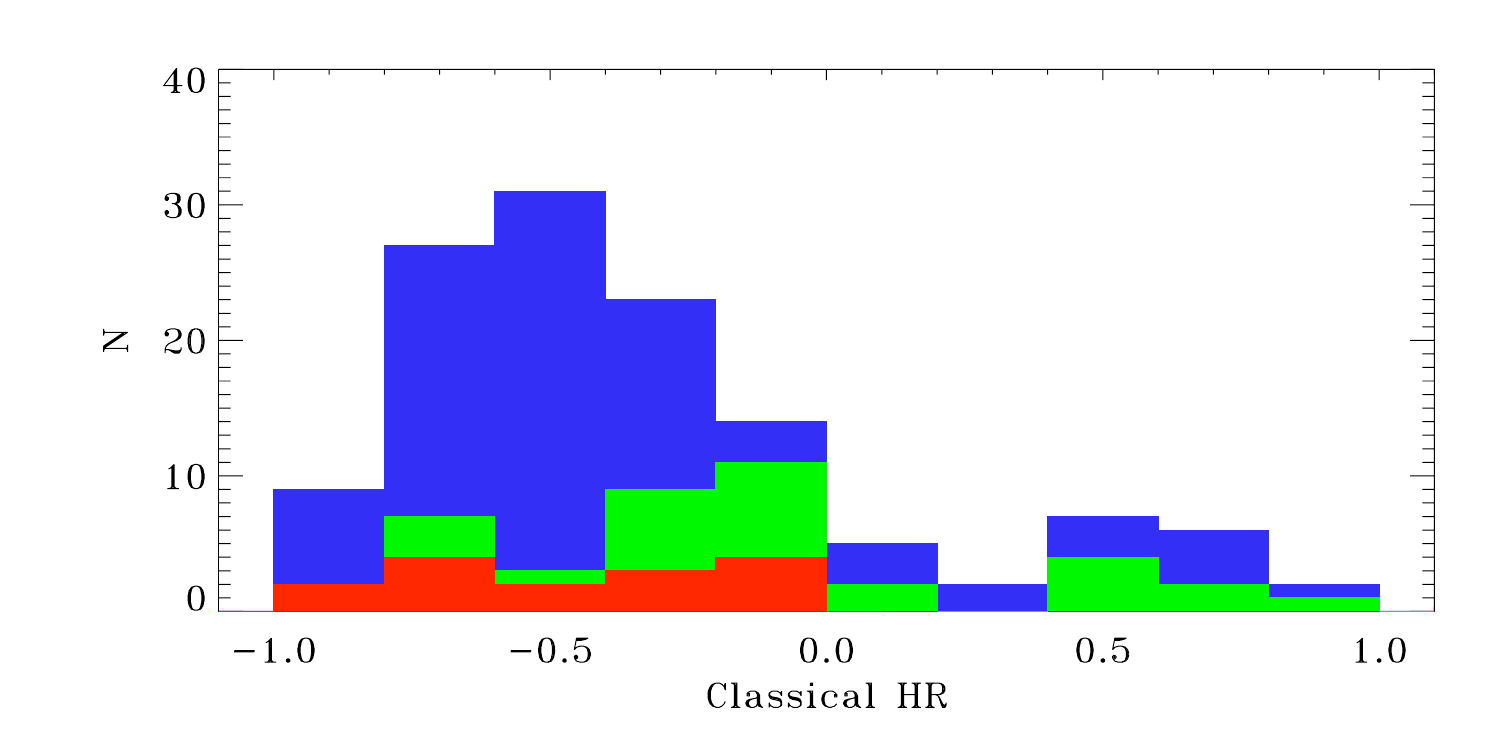}
 \caption{Histogram showing the distribution of the classical hardness ratio, using the soft band (0.5-2 keV) and the hard band (2-10 keV) photon fluxes from our sample of reference spectra. The blue histogram shows all sources, the green histogram shows \nh$>10^{23}$ \cmsq\ sources and the red histogram shows Compton thick sources.}
 \label{classichr_fig}
\end{figure}

In this paper we have aimed to identify an effective colour-colour selection for obscured AGN from X-ray data and improve upon the use of the classical hardness ratio. In Fig. \ref{classichr_fig} we plot histograms of the classical HR (with photon flux, rather than counts) for unabsorbed sources, absorbed Compton thin sources and Compton thick sources from \cite{brightman11}. This confirms our view that the hardness ratio is not an effective measure of obscuration, as all Compton thick sources appear in fact soft, while both Compton thin and unabsorbed sources show the full range of hardness ratio.

We have shown that colour-colour selection is able to break these degeneracies. The choice of bands for our colour-colour selection stems from the observation that while obscured AGN do appear hard above $\sim$4 keV, they are usually dominated by a soft excess component below 4 keV. This component is often scattered nuclear light, which is believed to have undergone Thompson scattering off hot electrons far from the nucleus, and hence out-with the obscuring material. It typically represents only a few percent of the intrinsic emission \citep{turner97}. Soft X-ray emission in obscured AGN can also be attributed to circumnuclear gas photoionised by the AGN \citep{guainazzi07}, or to star-formation related processes in the host galaxy, such as X-ray binaries. In all cases, this emission is assumed to be not subjected to the obscuration of AGN, and thus appears soft in HR1. 

{\bf \cite{dellaceca99} presented an analysis of hard X-ray selected sources from {\it ASCA} utilising very similar hardness ratios to ours. They also use these to infer absorption and spectral complexity in their sample. We have improved upon this technique to provide specific selection criteria for AGN with \nh$>10^{23}$ \cmsq\ as well as Compton thick AGN, and to  extend the analysis to higher redshifts}. Furthermore \cite{noguchi09} use similar hardness ratios in their study of the scattered fraction in AGN, also using data from 2XMM. They use the hardness ratios to select AGN with very small scattered fractions, in the search for buried AGN first described in \cite{ueda07}. They also note, however, that heavily obscured AGN can be selected well using these hardness ratios, though this is not the focus of their work.

Our results here provide a valuable technique for the identification of obscured AGN from photometry alone, when a spectral fit is not possible or badly constrained, and provides a direct method for investigating the nature of obscuration in AGN. As we have shown, it has applications for deep field, high redshift studies, where there are few good quality spectra, though above z$\sim$1, the 1-2 keV restframe band begins to be redshifted out of most X-ray telescope bandpasses. The redshift of the source should be known also, for correct application of the narrow bands. As \cite{noguchi09} suggested, it is well suited for application to wider and shallower surveys, such as the \xmm\ serendipitous source catalogue, especially as the photometry provided in that catalogue matches well to our choice of bands. {\bf While this method does very well at separating sources with \nh$>10^{23}$ \cmsq\ from sources with lower \nh, only X-ray spectroscopy can identify Compton thick sources, through the detection of the Fe K$\alpha$ line. This will be made possible for the low fluxes considered here with the high throughput X-ray spectral capabilities of {\it ATHENA}}

We have also used our hardness ratio analysis to make inferences about the spectral complexity CDFN sources {\bf as \cite{dellaceca99} did for {\it ASCA} sources}. We find that a significant number of CDFN sources at $z\sim$1 show spectral shapes more complicated than a simple absorbed power-law due to soft-excess emission. This has implications for the X-ray selection of heavily obscured sources, as it shows that they can be detected in the soft band through their soft-excess emission. This is good news for the {\it eRosita} instrument will conduct an all-sky imaging survey up to 10 keV. One of the primary science goals is to detect systematically all obscured accreting Black Holes in nearby galaxies and many (up to 3 million) new, distant active galactic nuclei. This technique provides a reliable tool for this purpose, especially if it is sensitive down to 0.2 keV, as this technique could be applied to sources out to $z\sim4$.

\section{Acknowledgements}

We thank the anonymous referee for their careful checking and critiquing of this manuscript which lead to its improvement. We also thank the builders and operators of {\it Chandra} and \xmm.

\bibliographystyle{mn2e}
\bibliography{bibdesk}


\label{lastpage}
\end{document}